\documentclass[12pt,epsf]{article}
\usepackage{epsfig}

\newcommand{\onefigure}[2]{\begin{figure}[htbp]
\begin{center}\leavevmode\epsfbox{#1.eps}\end{center}\caption{#2\label{#1}}
\end{figure}}

\newcommand{\twofigures}[3]{\begin{figure}[htdp]
\centering \leavevmode\epsfxsize=2.5in\epsfbox{#1.eps}
\leavevmode\epsfxsize=2.5in\epsfbox{#2.eps} 
\caption{{
#3}\label{#1}}
\end{figure}}

\renewcommand{\thanks}[1]{\footnote{#1}} 

\newcommand{\be}{\begin{equation}}
\newcommand{\ee}{\end{equation}}
\newcommand{\bea}{\begin{eqnarray}}
\newcommand{\eea}{\end{eqnarray}}

\begin{document}

\pagestyle{empty}

\bigskip\bigskip
\begin{center}
{\bf \large Self-consistent solutions of canonical proper self-gravitating 
quantum systems}
\end{center}

\begin{center}
James Lindesay\footnote{e-mail address, jlindesay@fac.howard.edu}\\
Computational Physics Laboratory \\
Howard University,
Washington, D.C. 20059 
\end{center}
\bigskip

\begin{center}
{\bf Abstract}
\end{center}

Generic self-gravitating quantum solutions that are
not critically dependent
on the specifics of microscopic interactions
are presented.
The solutions incorporate curvature effects,
are consistent with the universality of gravity,
and have appropriate correspondence
with Newtonian gravitation.
The results are consistent with known experimental
results that indicate the maintenance of the quantum
coherence of gravitating systems, as expected
through the equivalence principle.

\bigskip \bigskip \bigskip

\setcounter{equation}{0}
\section{Introduction}
\indent

The incorporation of quantum mechanics into
gravitational dynamics remains a perplexing
issue in modern physics. 
In contrast to other interactions like electromagnetism,
the trajectory of a gravitating system is independent
of the mass coupling to the gravitational field.  Thus
the gravitation of arbitrary test particles
can be described in terms of local geometry only.
However, the equations of general relativity are quite
complex and non-linear in the interrelations
between sources and geometry.  This makes solutions
of even classical systems somewhat complicated.

The behavior of quantum objects in Minkowski space-time
is well described by standard quantum theory.  The various
interpretations (Copenhagen, many worlds, etc.) of the underlying fundamentals of the
quantum world must all be consistent with the standard
theory, which has yet to be contradicted by experiment.
There have been experiments that examined the behaviors
of quantum objects in gravitational fields. 
Gravitating quantum systems \emph{do} maintain their
coherence, demonstrating that the structure of
the interaction need not break coherence in
order to localize the system in the field.
Moreover, those systems continue to gravitate after coherence
is broken by detection, as well as themselves serve as
source energy densities.  Even highly dynamic
gravitational environments such as the big bang
can redshift cosmic microwave background radiation
without breaking the coherence of the individual quanta.
These behaviors are sensible using fundamental principles of relativity. 
Due to the principle of equivalence, the motions of
detectors and screens should not break the coherence
of inertial (freely falling) systems prior to
detection.   These basics will be briefly discussed
in Section \ref{sec:QMandGravity}

Often, a problem considerably simplifies if the
parameters are properly chosen. 
In this treatment, space-like surfaces of
simultaneity will be defined by fixed proper
time $\tau$, which are generally not
coincident with space-like surfaces of simultaneity
defined by coordinate time $t$. 
The canonical proper time
formulation is particularly
useful for describing gravitational dynamics using proper time.
This formulation will be discussed in Section \ref{sec:ProperTimeGrav}.

Dynamics described using the proper time and convenient
spatial coordinates will provide straightforward solutions
for generic self-gravitating quantum systems.  The description will
be particularly robust with regards to arbitrary microscopic interactions
that might contribute to generate the inertial masses.  In fact, no form of
microscopic interaction is mentioned or utilized
in the discussion. The self-gravitating solutions
will be developed and discussed in Section \ref{sec:StaticSelfGrav}.

\setcounter{equation}{0}
\section{Quantum Mechanics and Gravity\label{sec:QMandGravity}}
\indent

Gravitation is an interaction of considerable mathematical
subtlety, despite its familiarity.  The geometrodynamics
of classical general relativity is most directly
expressed using localized geodesics.  However,
quantum dynamics incorporate measurement
constraints that disallow complete
localization of physical systems. 
The subtleties of observed gravitation of quantum systems
should offer insight into the fundamentals of
quantum self-gravitation, as will here be examined.

\subsection{Coherence of gravitating systems}
\indent

There is now considerable
experimental evidence that quantum coherence is maintained 
by the nearly static gravitational field near Earth's surface.  
During the early and mid 1970's, experiments performed by 
Overhauser, et.al.\cite{Overhauser} 
examined the gravitation of coherent 
neutrons diffracting from an apparatus whose orientation could 
be changed relative to the Earth's gravitational field.  
The gravitating neutrons were
seen to maintain spatial coherence, exhibiting a
pattern consistent with self-interference through
two apertures at different gravitational potentials.
A more recent experiment
measured the small difference between the ticks
of two interfering quantum clocks\cite{MullerPetersChu}.
In that experiment, very cold cesium atoms gravitated vertically across a laser
beam that superposed single atoms into states at differing gravitational
potentials.  The resultant difference in the phase demonstrated interference
in the quantum oscillations associated with the relativistic energy
of the atoms. 
Thus, gravitating atoms have also been shown to maintain temporal
coherence.

The experimental results discussed imply
that the quasi-static near-Earth
gravitational field does not break the phase coherence of 
neutrons or atoms as needed for interference.  The experiments
also test the principle of equivalence, since motions 
of the observer do not break the coherence of the inertial
(gravitating) particles.
The experiments involve both Newton's gravitational 
constant $G_N$ and Planck's constant $\hbar$  in a single equation 
form.

\setcounter{equation}{0}
\section{Canonical Proper Time Dynamics
\label{sec:ProperTimeGrav}}
\indent

Experiments such as those discussed in the previous section
provide illustrative examples of the usefulness of the
canonical proper time formalism developed in references
\cite{TGJL1993,JLTG2004}
for describing gravitating systems. 
The canonical proper time formulation
of relativistic dynamics provides a framework from which one 
can describe the dynamics of classical and quantum systems 
using the clocks of those very systems.  
The approach presumes that any
gravitating quantum system maintains coherence on surfaces defined
by \emph{its} proper time.  
The various regions across a coherent state
propagate through varying gravitational potentials,
with space-like surfaces of simultaneity defined by fixed proper time $\tau$.
The formulation utilizes a 
canonical transformation on the time variable
conjugate to the Hamiltonian that is used 
to describe the dynamics, but does not transform other dynamical 
variables such as momenta or positions.  
This gives
insight into the fundamentals of an 
interaction, since the response and back-response of the
interacting system is best parameterized using
its proper-scaled dynamics. 

\subsection{Proper time
Heisenberg equations}
\indent

For quantum systems, Heisenberg's equations describe the dynamics
of an observable $W(q,p,t)$ in terms of the commutator
of that observable with the Hamiltonian:
\be
{d \hat{W}(x,p,t) \over dt} = { i  \over \hbar}[\hat{H}, \hat{W}(x,p,t)] +
\left ( {\partial \hat{W} \over \partial t} \right )_{x,p}.
\ee
In special relativity,
the inertial time $t$ is related to the
proper time $\tau$ through the standard Lorentz 
factor\index{Lorentz factor} $\gamma$ using
\be
dt = \gamma \, d\tau = {H \over M c^2} d\tau.
\label{tgammatau}
\ee
The second form in Eqn. \ref{tgammatau} follows from the
relationship between the energy of a system compared to its rest
energy.  The canonical proper energy\index{canonical proper energy}
form $K$ is defined to generate dynamic changes with regards to the
\emph{proper time} of that system:
\be
{d\hat{W} \over d\tau} = {d\hat{W} \over dt}{dt \over d\tau} \equiv
{ i  \over  \hbar} [\hat{K},\hat{W}] + \left ( {\partial \hat{W} \over \partial \tau} \right )_{x,p}.
\ee
From this equation, along with Heisenberg's equation,
it then follows that $[K,W]={H \over M c^2}
[H,W]$. 

The canonical proper energy form $K$ is expected to correspond
to the Hamiltonian when the Hamiltonian itself corresponds to
the rest energy, $K|_{H=M c^2}=H=M c^2$. 
Holding the system mass $M$ fixed during the
canonical ``boost" from
$H$ to $K$ results in the form
\be
\hat{K}[H]={\hat{H}^2 \over 2 M c^2} + {M c^2 \over 2}.
\ee
As an example, a direct substitution of the non-interacting relativistic form
$H_o=\sqrt{(pc)^2+(M c^2)^2}$ into this equation yields
\be
\hat{K}_o={\hat{p}^2 \over 2 M} + M c^2 .
\ee
In this case, both temporal parameters are inertial.  
A few points of interest should be noted:
\begin{itemize}
\item The form of the equation for $K_o$ is that of a non-relativistic
free particle, despite the system being completely relativistic;
\item The momentum in $K_o$ is the same canonical momentum of the particle 
in the Hamiltonian formulation.  This is clearly \emph{not} a Lorentz
transformation of the dynamical parameters of the system;
\item The sometimes troublesome square root does not appear in
the expression for $K_o$.
\end{itemize}
Since the positions and momenta of a gravitating particle are typically
described relative to fiducial observers, rather than the proper coordinates
of the gravitating particle,
this formulation is particularly useful for describing gravitational
dynamics.

\subsection{Canonical proper time gravitation}
\indent

The equations of motion generated using the canonical proper time
formulation insures that the canonical proper energy is conserved
(${dK \over d\tau}=0$) if there is no explicit temporal dependence
in the functional form of any interactions. 
For generic proper potential energy forms
$U(\mathbf{r})$, the canonical proper energy can often be expressed
\be
K~=~ {\mathbf{p} \cdot \mathbf{p} \over 2 m} + U(\mathbf{r}) + m c^2 .
\label{GenKeqn}
\ee
A potential form consistent with standard gravitation will next be
developed.

\subsubsection{Curvature effects}
\indent

The potential energy is expected to take the form of Newtonian
gravitation to lowest order in the gravitational
constant $G_N$.  However, space-time curvature
effects are expected to modify the classical result. 
To construct the relativistic energy form, the
equations of motion resulting from Eqn. \ref{GenKeqn}
with $U(\mathbf{r})=m \, V(\mathbf{r})$ will be examined:
\be
{d p_j \over d \tau} ~=~ - m \,  \partial_j V(\mathbf{r}) \quad , \quad
{d r_j \over d \tau} ~=~ {p_j \over m} \quad .
\ee
For a constant mass, this form is analogous to the
geodesic equation incorporating space-time curvature:
\be
{d^2 x^j \over d \tau ^2} + 
\Gamma_{\alpha \beta}^j  {d x^\alpha \over d \tau}{d x^\beta \over d \tau}~=~0 .
\label{GeodesicEqn}
\ee
The present exploration is interested in the behaviors of
stationary quantum gravitating systems. 
As is the case with electronic distributions in stable
atoms, the mass distribution should be stationary in
a quantum gravitating system. 
For a stationary gravitating distribution, assume that
${d x^\alpha \over d \tau}={d x^0 \over d \tau} \delta_0 ^\alpha$
(consistent with quantum expectation values).
Therefore, substituting the form of the connections $\Gamma_{\alpha \beta}^j $
for a metric space-time (Riemannian manifold)
in the geodesic equation \ref{GeodesicEqn}, the proper
interaction form must satisfy
\be
{d^2 x^j \over d \tau ^2} ~=~ - \partial_j V(\mathbf{r}) ~=~
{1 \over 2} g^{j \mu} \, g_{00, \mu}  \left ( {d x^0 \over d \tau}  \right )^2.
\label{MetricDynamics}
\ee
This form will be generated for a straightforward static
energy density.

\subsubsection{Form of the metric \label{subsec:metric}}
\indent

The space-time metric for a  spherically
symmetric, static space-time will be chosen to be a
generalization of Schwarzschild geometry with non-vanishing
local densities.  The metric is given by
\be
ds^2 ~=~ -\left ( 1 - {R_M (r) \over r}  \right ) \, (dct)^2 + 
{ dr^2 \over 1 - {R_M (r) \over r}  } + r^2 d\theta^2 +
r^2 sin^2 \theta \, d\phi^2 ,
\label{metric}
\ee
where the Jacobian $\sqrt{-det \, \mathbf{g}}
\equiv \sqrt{-g}=r^2 sin\theta$.
In this equation, a finite radial mass scale
$R_M (r) \equiv 2 G_N M(r) / c^2$ is the length scale of the
interior mass-energy content of the system,
with the mixed Einstein tensor given by
${G^0}_0~=~ {1 \over r^2} {\partial \over \partial r}R_M (r)$.
For finite mass distributions, the metric takes the
form of Minkowski space-time
both asymptotically ($r >> R_M $) as well as wherever the
radial mass scale vanishes. 
The Ricci scalar
\be
\mathcal{R} ~=~ -{1 \over r^3} {d \over dr}(r^2 \, {d R_M(r) \over dr})
\ee
for such distributions
is non-singular as long as the mass density decreases rapidly enough for
small $r$.

\subsubsection{Proper potential form}
\indent

Substituting the metric into Eqn. \ref{MetricDynamics},
one should note that $g^{rr} = - g_{00}$ and
$ \left ( {d x^0 \over d \tau}  \right )^2=-{c^2 \over g_{00}}$, giving the equation
\be
\partial_r \, V(r) ~=~ -{c^2 \over 2}
\partial_r \, (g_{00})
\label{PotentialMetric}
\ee
Using the standard condition $V(\infty)=0$, the form of the
interaction for the proper canonical energy form is
therefore given by
\be
V(r)~=~
- {G_N M(r) \over r} .
\label{PotentialEnergy}
\ee
This relativistic form is the same as
the usual Newtonian interaction.

\setcounter{equation}{0}
\section{Static field quantum self-gravitation
\label{sec:StaticSelfGrav}}
\indent

The gravitational potential energy from Eqn. \ref{PotentialEnergy}
will next be incorporated in the quantum form of the canonical
proper energy equation \ref{GenKeqn}.  The equation developed
will include both the local dynamics of special relativity as well
as the curvature effects of general relativity.  Self-consistent solutions
of this equation will be developed in this section.

\subsection{Proper time quantum gravitating particles}
\indent

Consider the stationary gravitation of
a mass $m$ due to an interior source mass distribution $M(r)$. 
An invariant probability form measuring the likelihood that the
particle will be measured by an observer in the space-time interval
$\Delta ct \, \Delta \mathcal{V}$ is expected to take the form
\be
\mathcal{P}_{\Delta ct \, \Delta \mathcal{V}}~=~
\int_{\Delta ct \, \Delta \mathcal{V}} \, dct \, d^3 r \,  \sqrt{-g} \,
|\psi(ct,\mathbf{r})|^2 .
\ee
General quantum systems will be temporally dynamic. 
However, stationary state probability densities are not
expected to have time dependencies.
The wave function that satisfies the stationary
state canonical proper energy
equation for this mass,
and represents the likelihood for measurement
within the time interval $\Delta ct$,
is given by
\be
\begin{array}{c}
\left [ {\hat{\mathbf{p}} \cdot \hat{\mathbf{p}} \over 2 m } - 
{G_N m M_\ell (r) \over r} + 
m  c^2 \right ]
\psi_{n \ell \ell_z}^{\Delta ct} (r,\theta,\phi) ~=~ K_{n \ell} ~ \psi_{n \ell \ell_z}^{\Delta ct} (r,\theta,\phi)
\quad , \\ \\
\psi_{n \ell \ell_z}^{\Delta ct} (r,\theta,\phi)={1 \over \sqrt{\Delta ct}}
R_{n \ell} (r) Y_\ell ^{\ell_z} (\theta, \phi) .
\end{array}
\label{mgravGenWF}
\ee
The proper energy eigenvalues $K_{n \ell}$ are expected to include relativistic velocities
and temporal curvature effects.

The form of the canonically conjugate momentum components in Eqn. \ref{mgravGenWF}
must be consistent with the Heisenberg equations of motion
\be
\left \langle {d \hat{p}_r \over d\tau} \right \rangle ~=~
\left \langle  {i \over \hbar}[\hat{K},\hat{p}_r]  \right \rangle ~=~
-m \,  \partial_r V(r) ,
\ee
where $V(r)$ was obtained from the geodesic equation \ref{PotentialMetric}. 
This implies that the scale factor of the momentum conjugate to $r$ in
the proper energy form should be unity, $\hat{p}_r = {\hbar \over i}{\partial \over \partial r}$.
Therefore, the spatial curvature effects are evidently already incorporated in the
functional form of the potential $V(r)$ and the given conjugate momentum operator.

The square of the momentum for the metric form Eqn. \ref{metric} is thus given by
\be
\hat{\mathbf{p}} \cdot \hat{\mathbf{p}} ~=~ - \hbar ^2 \left \{
{1 \over r^2}  {\partial \over \partial r} 
\left ( r^2  {\partial \over \partial r}  \right ) -
{\hat{\mathbf{L}}^2 \over \hbar^2 r^2}
\right \} ,
\ee
while in contrast the spatial Laplacian for this metric satisfies
\be
\mathbf{\nabla} \cdot \mathbf{\nabla} ~=~  \left \{
{1 \over r^2} \sqrt{1 - {R_m (r) \over r}} {\partial \over \partial r} 
\left ( r^2 \sqrt{1 - {R_m (r) \over r}} {\partial \over \partial r}  \right ) -
{\hat{\mathbf{L}}^2 \over \hbar^2 r^2}
\right \} .
\ee
Parameters analogous to those of Bohr for hydrogenic
systems can be developed.  The radial scale of the solutions
is given by
\be
a ~\equiv~ {\hbar^2 \over G_N m^3} ~=~
\left ( {\lambda_m  \over L_P } \right )^2 \lambda_m
~=~ \left ( {M_P \over m} \right )^2 \lambda_m ,
\label{mBohrRadius}
\ee
where the reduced Compton wavelength
is given by $\lambda_m  \equiv {\hbar \over m c}$,
the Planck length is labeled $L_P$, and the Planck mass is labeled $M_P$.
For the present treatment, only s-wave $\ell=0$ states will be examined. 
Since probability densities using the metric
coordinates will take the form $ r^2 R_{C,\ell=0}^2 (r)$,
It is convenient to introduce a central reduced radial wavefunction $u_C (r/a)\propto
r R_{C,0} (r)$ parameterized by dimensionless variable $\zeta \equiv r/a$.
The dynamic parameters can also be scaled using the parameter $a$:  
\be
\begin{array}{l}
\mathcal{P}(r/a) ~\equiv~ \int_0^{r/a} \, u_{C}^2 (\zeta') \,  d \zeta' ~,~ \mathcal{P}(\infty)=1 ~,\\ \\
{R_M (r) \over r} ~=~ 2 \, \left ( {m \over M_P}  \right )^4 \, {a \over r} \, \mathcal{P}(r/a) ~, \\ \\
-{2 m a^2 \over \hbar^2} V(r) ~=~ 2 \, { a \over r}\, { M(r/a) \over m} ~, \\ \\
-{2 m a^2 \over \hbar^2} (K - m c^2) ~=~ -2 \left(
{M_P \over m} \right )^4  {K - m c^2 \over m \, c^2} ~.
\end{array}
\label{rescaling}
\ee

Using these identifications,
Eqn. \ref{mgravGenWF} can then be re-written
\be
{\epsilon_C} \,  u_C (\zeta) ~=~ {d^2 u_C (\zeta) \over d \zeta^2 } ~+~
\left ( {2 \over \zeta} \right ) \left (  M (\zeta) \over m
\right )  u_C (\zeta)  ~ ,
\label{uGenSEqn}
\ee
where the dimensionless parameter $\epsilon_C \equiv -2 \left(
{M_P \over m} \right )^4  {K - m c^2 \over m \, c^2}$
quantifies the gravitational
binding energy of the mass.  In order to examine the scale
of the spatial curvature effects, an equation for which the
spatial Laplacian (which incorporates proper radial distances)
replaces $-{\hat{\mathbf{p}} \cdot \hat{\mathbf{p}} \over \hbar^2}$, will
also be examined:
\be
\begin{array}{l}
{\epsilon_*} \,  u_* (\zeta) ~=~ {d^2 u_* (\zeta) \over d \zeta^2 } ~+~
\left ( {2 \over \zeta} \right ) \left (  M_* (\zeta) \over m
\right )  u_* (\zeta) ~ +  ~ \\ 
\left ( {m \over M_P} \right )^4 \left (
\left [ {\mathcal{P}'_*(\zeta) \over \zeta^2} -
  {\mathcal{P}_*(\zeta) \over \zeta^3}  \right ]  u_* (\zeta) + 
\left [   {\mathcal{P}_*(\zeta) \over \zeta^2} -
  {\mathcal{P}'_*(\zeta) \over \zeta} 
\right ] {d u_* (\zeta) \over d \zeta } -
{2 \over \zeta} \mathcal{P}_*(\zeta) \, {d^2 u_* (\zeta) \over d \zeta^2 } \right )  .
\end{array}
\label{uGenSEqnRm}
\ee
The terms containing
$\mathcal{P}_*(r)$ demonstrate the modifications to the previous form
(with ${m \over M_P} \rightarrow 0$) due to factors ${R_{M_*} \over r}$.

For all solutions,
the source mass $M (\zeta)$ will be presumed
to be generated by the interior self-sourcing  of probability density:
\be
M(\zeta) = \int_0^\zeta \rho_{mass} (\zeta') \, d\zeta' =
\int_0^\zeta 
 \, m \, u_C ^2 (\zeta') \,
d \zeta' = m \mathcal{P}(\zeta) .
\label{MassProb}
\ee
The distribution indicates that the differential equations
\ref{uGenSEqn} and \ref{uGenSEqnRm} will be
non-linear.  For this distribution, it should be noted that the particle mass scale $m$
appears nowhere in Eqn. \ref{uGenSEqn}, while it only appears
in the spatial scale terms in Eqn. \ref{uGenSEqnRm}.

A mass whose interior probability density provides its local source gravitational
field will be referred to as a \emph{self-gravitating}\index{self-gravitation} mass.  If,
in additional, the overall gravitational bound state energy is that
of the mass itself, the self-gravitating mass
will be referred to as a \emph{self-generating}\index{self-generation} mass.
A self-gravitating central mass is expected to have
non-vanishing probability density at the center. 
Such a self-gravitating single mass satisfies Eqn. \ref{uGenSEqn} with
mass distribution given in Eqn. \ref{MassProb}. 
This form is clearly non-linear, so that initial conditions and
eigenvalues are non-trivially related to the solution. 

A solution to Eqn. \ref{uGenSEqn} for
a system that is self-gravitating, but has non-vanishing binding energy eigenvalue
is demonstrated in Figure \ref{SGruC}.
\twofigures{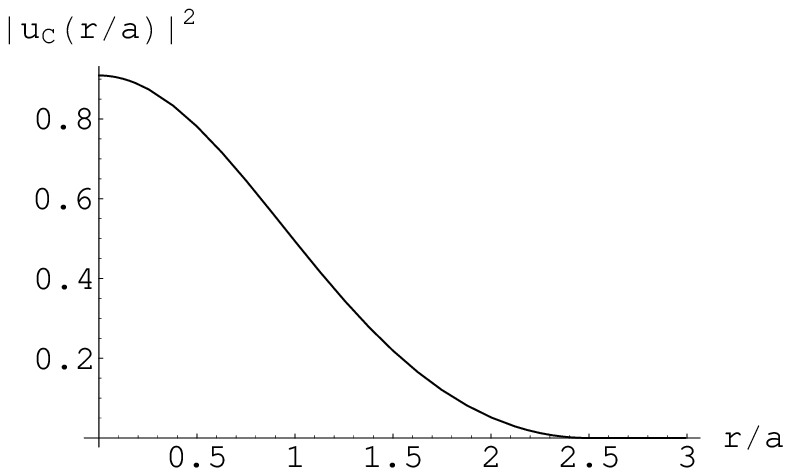}{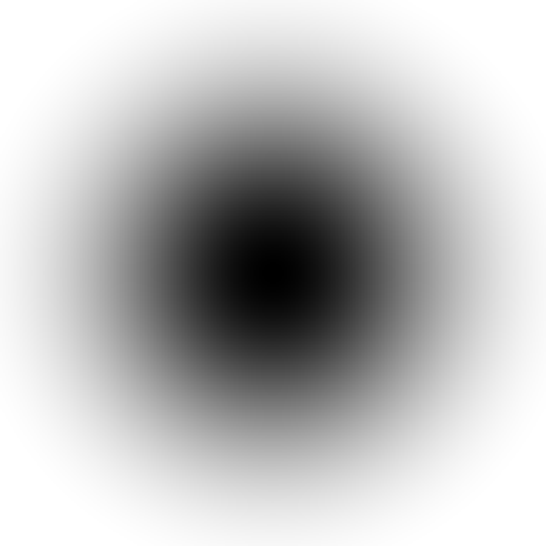}{Self gravitating quantum mass distribution.}
Expressed in the dimensionless form demonstrated, the solution is completely
independent of the mass of the system. 
The binding energy eigenvalue for the normalized probability density was obtained by examining
the small $r$ behavior of Eqn. \ref{uGenSEqn}
in a self-consistent manner, yielding a value
$\epsilon_C \simeq 1.212$. 
The diagram on the left demonstrates the probability density
$|u_C (r/a)|^2$, while the diagram on the right is a density plot
of the self-gravitating mass density.
For the system, the gravity at a
given radial coordinate is a field generated by the
integrated mass density within that radial coordinate.  

To obtain a solution to  Eqn. \ref{uGenSEqnRm}, a mass
value must be chosen, since the radial scale factors explicitly appear in the equation. 
The mass was chosen ($\left ( {m \over M_P}  \right )^4=0.01$, or
$m \simeq 0.316 \, M_P$)
such that \emph{some} spatial curvature effects would be apparent
in the calculations.  For this mass,
the binding energy eigenvalue $\epsilon_*$ was found to be
larger than $\epsilon_C$
by  0.38\%, while
the central density remained essentially unchanged.  

A solution  to Eqn. \ref{uGenSEqn} for which the gravitational
potential results in vanishing net binding energy can also be found. 
Figure \ref{SGnuC} demonstrates a self-generating
solution to this equation with $\epsilon_C=0$.
\twofigures{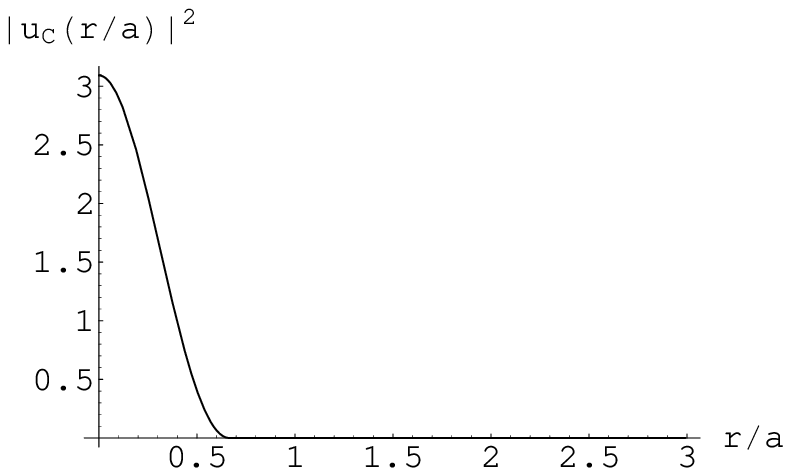}{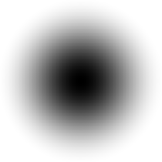}{Self generating quantum mass distribution.}
The diagram on the left again demonstrates the probability density
$|u_C (r/a)|^2$, while the diagram on the right is a density plot
of the self-gravitating mass density consistent with vanishing
overall gravitational binding energy.  The scales of the diagrams
have been chosen to be consistent with the prior self-gravitating
mass.  The self-generating mass density is seen to
be more concentrated at the center relative to the self-gravitating mass density,
developing a greater integrated gravitational potential energy, with a commensurate
change in integrated kinetic energy.  The central density $|u_* (0)|^2$
solving Eqn. \ref{uGenSEqnRm}  (with $m \simeq 0.316 \, M_P$) is modified from that
solving Eqn. \ref{uGenSEqn} by an increase of about
3.8\%.

The solutions demonstrated in Figures \ref{SGruC} and \ref{SGnuC}
are independent of mass. However, as previously mentioned, the radial mass scale used to
generate the Einstein tensor is given by 
$R_M (r)  ~=~ 2 \, \left ( {m \over M_P}  \right )^4 \, a \, \mathcal{P}(r/a) $.
The crucial factor $1-{R_M (r) \over r}$ in the metric is demonstrated in Figure \ref{SGnRm}.
\onefigure{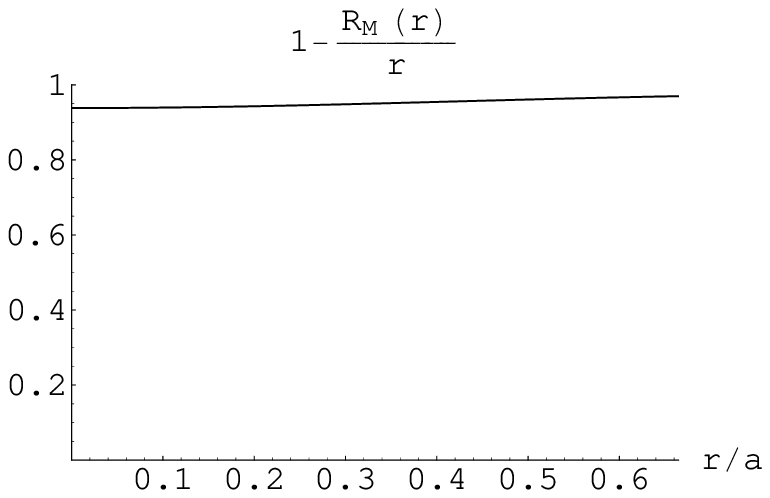}{Metric factor $-g_{00}$ and $g_{rr}^{-1}$ .}
If this factor changes sign, space-like behaviors become time-like (and vice versa), and a trapped
region for which outgoing photons must propagate towards decreasing radial parameter $r$
will be present.  As long as the mass is not chosen to be too large,
there is no trapped region. 
A larger mass lowers the y-intercept of this curve.
 For the self-generating mass, the maximum value the mass can take without introducing
a trapped region is given by about 0.63 $M_P$.  Masses smaller than this will not
generate a black hole.

For completeness, the non-vanishing components of the
Einstein tensor ${G^{ct}}_{ct}={G^{r}}_{r}$ and
${G^{\theta}}_{\theta}={G^{\phi}}_{\phi}$ are
demonstrated in Figure \ref{SGnG00}.
\twofigures{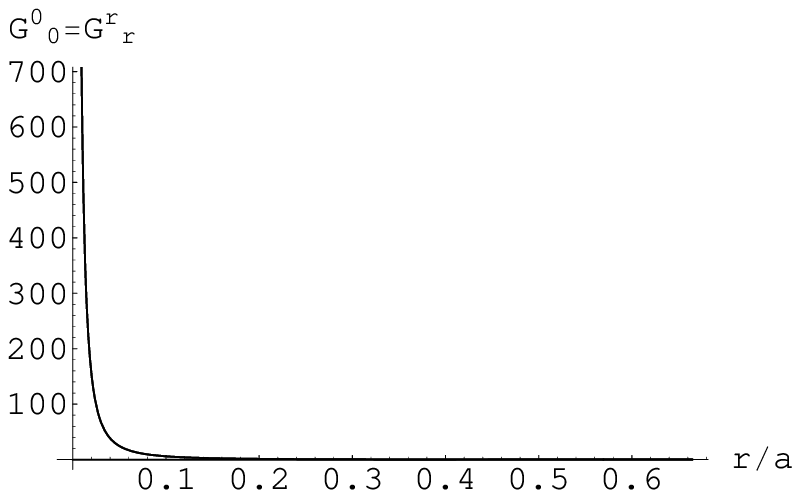}{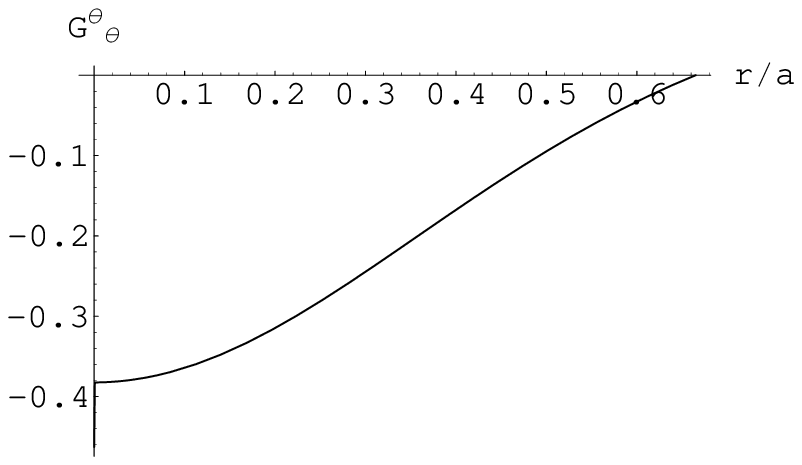}{Einstein tensor components
for self-generating mass.}
The tensor satisfies the vacuum solution ${G^{\mu}}_{\beta}=0$
in the exterior region ${r \over a}>0.64$.  All solutions presented
have non-vanishing densities of finite extent.

\subsection{Energy conditions}
\indent

Classical gravitating systems are expected to satisfy various energy
conditions everywhere.  These conditions assert that any observer
should locally measure gravitational fields generated by
time-like or light-like sources, regardless of their motion. 
This is consistent with an expectation that no energy source
can propagate at a speed greater than that of light.  However,
quantum systems \emph{do} exhibit space-like coherent
behaviors. Space-like coherence allows the evaporation of
black holes, thereby locally violating energy conditions. 
Also, systems with significant binding might
violate these conditions.  It is therefore of interest to examine the
energy conditions of these self-gravitating systems.

The \emph{null} and \emph{weak} energy conditions
assert that the form
\begin{displaymath}
\mathcal{I}_{null/weak}
 \equiv -u_{null/weak}^\mu \, T_{\mu \beta} \, u_{null/weak}^\beta
\le 0 
\end{displaymath}
should be non-positive for light-like (null) and time-like (weak) observer
four velocities, where $T_{\mu \beta}$ represents components of the
energy-momentum tensor sourcing the gravitational field in Einstein's equation. 
The \emph{dominant} energy condition directly develops the
form of the 4-momentum of the gravitational source as seen
by the observer with four velocity $\vec{u}_{observer}$,
given by
$p_{source}^\mu \equiv -{T^\mu}_\beta \, u_{observer}^\beta$. 
This four-momentum is expected to be time-like or light-like, i.e.,
\begin{displaymath}
\mathcal{I}_{observer}^{DE} \equiv \vec{p}_{source} \cdot \vec{p}_{source} \le 0,
\end{displaymath}
where the dot product is defined by the metric of the geometry.
For all of the self-gravitating \emph{and} the self-gravitating, self-generating
solutions given, the
null and weak energy conditions are satisfied everywhere for all types of motions.
Likewise, the dominant energy condition for fiducial (static) observers and
arbitrary radial motions is also satisfied everywhere 
for all solutions.
However, the dominant energy condition for rapid pure azimuthal motions
of the observer
was found to be violated only in the region
just inside of the surface in each solution, likely due to
coherence and gravitational binding from the interior mass
distribution.  
Rapid motions were found to be those motions exceeding the condition
\be
r \, u_\theta ~ > ~ {2 R_M '(r) \over \sqrt{(r \, R_M ''(r))^2 - (2 R_M '(r))^2}}.
\ee
Exterior to the region of coherence, as well as proximal to
the center, all energy conditions were found to be satisfied. 
A further exposition of energy conditions, as well as a more
detailed development of the general formulation based on the
equivalence principle, including co-gravitating
masses and cluster decomposability, will be found in reference \cite{JLCUPress}.

\setcounter{equation}{0}
\section{Conclusions}
\indent

Self-gravitating quantum solutions, consistent
with the equivalence principle, have been found using coherence
parameterized by the local proper time of the gravitating system.
The solutions required no specific form for micro-physical interactions,
consistent with the universal nature of gravitation.
The approach considers \textit{space-time} as an emergent construct
of quantum measurement, with curvatures generated by
Einstein's equation in the form
$G_{\mu \nu}=- {8 \pi G_N \over c^4} \langle \hat{T}_{\mu \nu} \rangle$. 
The dynamics developed is consistent with the measurement constraints
of standard quantum theory.

The quantum stationary solutions
developed incorporate curvature effects.  For weak
curvatures and slow motions, the solutions exhibit
both quantum and classical Newtonian correspondence
through proper time Heisenberg equations of motion.
The formulation, being generally representation
independent, demonstrates that the exhibition of quantum coherent 
behavior for gravitating systems need not require second quantization 
of the gravitation field itself.  
The solutions satisfy sensible conditions of physicality on the energy
densities sourcing the gravitational fields,
including non-singular behavior everywhere and
non-negative mass densities $R_M ' (r) \ge 0$ everywhere.
The natural size scales are consistent with bound quantum systems.

Interaction forms for more general metric spaces have also been developed. 
The approach is consistent with experimental results demonstrating the
coherence of gravitating systems.
Because of the non-linearity of the equations, (independent) co-gravitating masses
modify the solutions as expected.
These results, as well as a more detailed analysis of its general
foundations, will be presented elsewhere\cite{JLCUPress}.

\bigskip
\begin{center}
\textbf{Acknowledgments}
\end{center}
The author gratefully acknowledges useful past discussions with 
Tepper Gill, Tehani Finch, and Lenny Susskind.

\end{document}